\documentclass{article}

\usepackage{arxiv}

\usepackage[utf8]{inputenc} 
\usepackage[T1]{fontenc}    
\usepackage{hyperref}       
\usepackage{url}            
\usepackage{booktabs}       
\usepackage{amsfonts}       
\usepackage{nicefrac}       
\usepackage{microtype} 
\usepackage{caption}
\usepackage{subcaption}
\usepackage{graphicx}

\usepackage{stackrel}
\title{Application of deep reinforcement learning for Indian stock trading automation}

\author{
  Supriya~Bajpai \\
  IITB-Monash Research Academy, IIT Bombay, India\\
  Monash University, Australia\\
  \texttt{supriya.bajpai@monash.edu} 
}

\begin{document}
\maketitle

\begin{abstract}
In stock trading, feature extraction and trading strategy design are the two important tasks to achieve long-term benefits using machine learning techniques. Several methods have been proposed to design trading strategy by acquiring trading signals to maximize the rewards. In the present paper the theory of deep reinforcement learning is applied for stock trading strategy and investment decisions to Indian markets. The experiments are performed systematically with three classical Deep Reinforcement Learning models Deep Q-Network, Double Deep Q-Network and Dueling Double Deep Q-Netwonrk on ten Indian stock datasets. The performance of the models are evaluated and comparison is made.
\end{abstract}


\section{Introduction}
A lot of work has been done to propose methods and algorithms to predict stock prices and optimal decision making in trading.  A large number of indicators, machine learning and deep learning techniques~\cite{hiransha2018nse} such as Moving averages~\cite{fifield2008performance, ariyo2014stock}, linear regression~\cite{bhuriya2017stock,cakra2015stock,trafton2011memory}, neural networks~\cite{dutta2006artificial,ahangar2010comparison}, Recurrent neural network~\cite{berradi2019integration, kim2019forecasting, nelson2017stock, moghar2020stock} and Reinforcement learning (RL) have been developed to predict the stock and financial price and strategies~\cite{pendharkar2018trading,meng2019reinforcement}. The advanced techniques of artificial
neural network have shown better performance as compared to the traditional indicators and methods~\cite{kuo1998decision,baba1992intelligent}. The stock price prediction is a very challenging task as the stock market changes rapidly and data availability is also incomplete and not sufficient. Reinforcement learning is one of the methods to solve such complex decision problems. Reinforcement learning can prove to be a better altemative approach for stock price prediction~\cite{lee2001stock} and maximizing expected return. Deep Learning methods have the ability to extract features from high dimentional data. However, it lacks the decision-making capabilities. Deep Reinforcement Learning (DRL) combines the Deep Learning approach with the decision making ability of Reinforcement Learning.  Researchers have investigated RL techniques to solve the algorithmic trading problem. Recurrent Reinforcement Learning (RRL) algorithm have been used for discovering new investment policies without the need to build forecasting models~\cite{moody2001learning}.
Adaptive Reinforcement Learning (ARL) have been used to trade in foreign exchange markets~\cite{dempster2006automated}. Recently, people investigated DRL method to solve the algorithmic trading problem~\cite{li2019application, xiong2018practical, deng2016deep,carapucco2018reinforcement,boukas2020deep, li2019application, lee2019global}. 

In the present paper Deep Reinforcement Learning is applied to Indian stock market on ten randomly selected datsets to automate the stock trading and to maximize the profit. Model is trained with historical stock data to predict the stock trading strategy by using Deep Q-Network (DQN), Double Deep Q-Network (DDQNn) and Dueling Double Deep Q-Network (Dueling DDQN) for holding, buying and selling the stocks. The model is validated on unseen data from the later period and performance is evaluated and compared.

\section{Methods}
Deep Q-Network, Double Deep Q-Network and Dueling Double Deep Q-Network~\cite{sewak2019deep} are discussed in the following sections.
\subsection{Deep Q-Network}
\begin{figure}[h!]
\includegraphics[width=\linewidth]{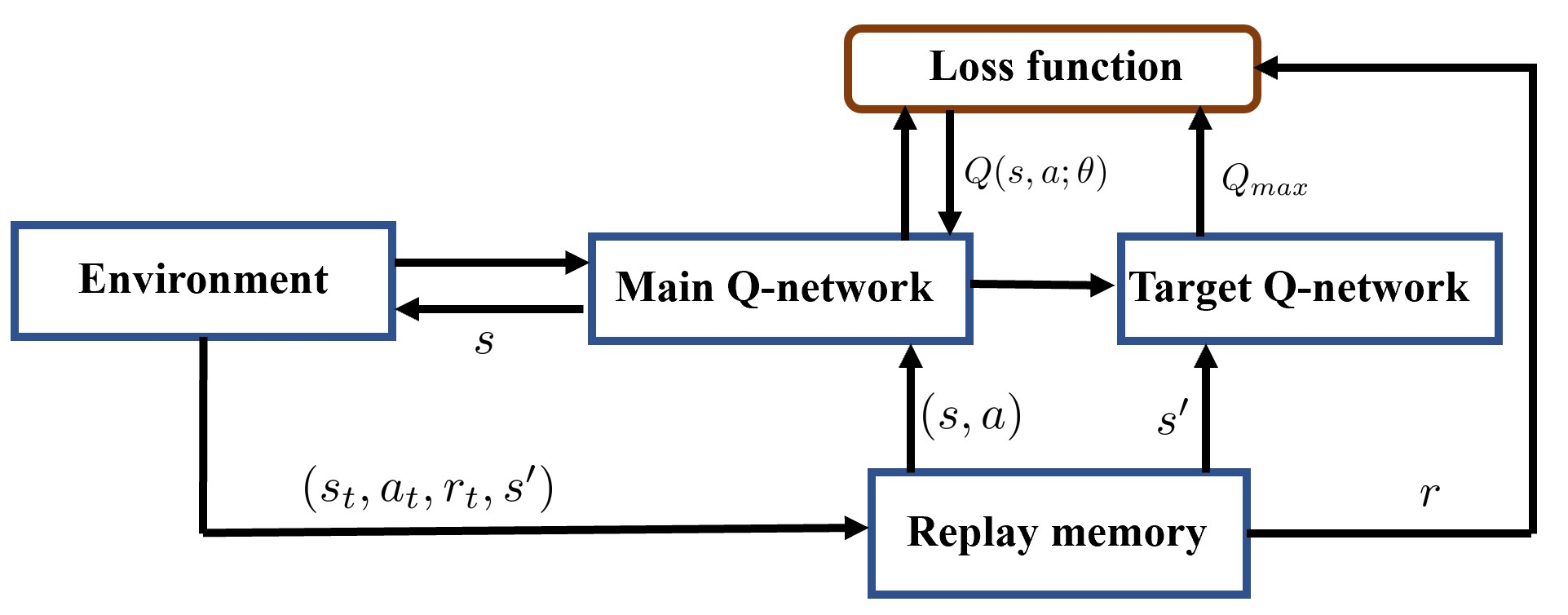}

\caption{Deep Q-Network model architecture.}
\label{fig:1}
\end{figure}

Deep Q-Network is a classical and outstanding algorithm of Deep Reinforcement Learning and it's model architecture is shown in Figure~\ref{fig:1}. It is a model-free reinforcement learning that can deal with sequential decision tasks. The goal of the learning is to learn an optimal policy $\pi^{\star}$ that maximizes the long term reward or profit. The agent takes action $a_t$ depending on the current state $s_t$ of the environment  and receives reward $r_t$ from the environment. The experience replay is used to learn from the previous experiences and is used to store the previous states, actions, rewards, and next states. The data from the replay memory is sampled randomly and fed to the train network in small batch sizes to avoid overfitting. In deep Q-learning the Convolutional Neural Network (known as Q-Network) is used to learn the expected future reward Q-value function ($Q(s_t,a_t)$). One major difference between the Deep Q-Network and the basic Q-learning algorithm is a new Target-Q-Network, which is given by:
\begin{equation}
    Q_{target}=r_{t+1}+\gamma max_{a'} [Q(s'_t,a'_t;\theta)]
\end{equation}
where, $Q_{target}$ is the target Q value obtained using the Bellman Equation and $\theta$ denotes the parameters of the Q-Network. In DQN there are two Q-Networks: main Q-Network and target Q-Network. The target Q-Network is different from the main Q-Network which is being updated at every step. The network values of the target Q-Network are the updated periodically and are the copy of the main network’s values. Use of only one Q-Network in the model leads to
delayed or sub-optimal convergence when the data incoming frequency is very high and the training data is highly correlated and it may also lead to unstable target function. The use of two different Q-Networks increases the stability of the Q-Network.

Optimal Q-value or the action-value pair is computed to select and measure the actions. DQN takes the max of all the actions that leads to overestimation of the Q-value, as with the number of iterations the errors keeps on accumulating~\cite{van2016deep}. This problem of overestimation of Q-value is solved by using Double DQN, as it uses another neural network that optimizes the influence of error.  

\subsection{Double Deep Q-Network}
The above problem of overestimation becomes more serious if the actions are taken on the basis of a Target Q-Network as the values of the Target Q-Network are not frequently updated. Double DQN uses two neural networks with same structure as in DQN, the main network and the target network as it provides more stability to the target values for update. In Double DQN the action is selected on the basis of the main Q-Network but uses the target state-action value that
corresponds to that particular state-action from the Target Q-Network. Thus, at each step all the action-value pairs for
all possible actions in the present state is taken from the main Q-Network which is updated at each time step. Then an argmax is taken over all the state-action
values of such possible actions~(Equation~\ref{eq:2}), and the state-action value which
maximizes the value, that specific action is selected.
\begin{equation}\label{eq:2}
    Q_{target}=r_{t+1}+\gamma Q(s_t, argmax_{a'} Q(s_{t}^{'},a_{t}^{'}; \theta); \theta^{'})
\end{equation}
But to update the main Q-Network the value that corresponds to the selected state-action pair
is taken from the target Q-Network. As such we can overcome both the problems of overestimation and instability in Q-values.
\subsection{Dueling Double Deep Q-Network}
There are two Q-Networks in both DQN as well as in Double DQN, one is the main network and the other is the target network where the network values are the periodic copy of the main network’s values. The Dueling Double DQN has non-sequential network architecture where, the convolutional layers get separated into two streams and both the sub-networks have fully connected layer and output layers. The first sub-network corresponds to the value function to estimate the value of the given state and the second sub-network estimates the advantage value of taking a particular action over the base value of being in the current state.
\begin{equation}\label{eq:5}
    Q(s_t,a_t;\theta,\alpha,\beta)=V(s_t;\theta,\beta)+(A(s_t,a_t;\theta,\alpha)-\stackrel[a'\in |A|]{}\max A(s_t,a'_{t};\theta,\alpha))
\end{equation}

here, $A$ is the advantage value. We can get the Q-values or the action-value by combining the output of the first sub-network, that is the base value of state with the advantage values of the actions of the second sub-network. $\theta$ is common parameter vector both the sub-networks. $\alpha$ and $\beta$ are the parameter vectors of the “Advantage” sub-network and State-Value function respectively. The Q value for a given state-action pair is equal to the value of that state which is estimated from the state-value $(V)$ plus the advantage of taking that action in that state. We can write the above Equation~\ref{eq:5} as follows. 
\begin{equation}\label{eq:6}
    Q(s_t,a_t;\theta,\alpha,\beta)=V(s_t;\theta,\beta)+(A(s_t,a_t;\theta,\alpha))
\end{equation}
From the above equation we can get the Q-value if we know the value of S and A, but we cannot get the values of S and A if Q-value is known. The last part of the Equation~\ref{eq:5} is slightly modified as follows, which also increases the stability of the algorithm.

\begin{equation}\label{eq:7}
    Q(s_t,a_t;\theta,\alpha,\beta)=V(s;\theta,\beta)+(A(s_t,a_t;\theta,\alpha)-\frac{1}{|A|}\sum_{a'}A(s_t,a'_t;\theta,\alpha))
\end{equation}

\section{Experiments}
In the present study we evaluate the performance of the  deep reinforcement learning algorithms for  stock market investment decisions on 10 Indian stock dataset. The dataset is obtained from National Stock Exchange (NSE) India, that consists of the price history and trading volumes of stocks in the index NIFTY 50. We used Deep Q-Network (DQN), Double Deep Q-Network (DDQN), and Dueling Double Deep Q-Network (Dueling DDQN)  to  automate  the  stock  trading  and  to  maximize  the profit. We split the dataset for training and testing purpose in equal proportions. The training and testing dataset is fed to the models and the train and test rewards and profit are estimated and compared.

\begin{figure}[h!]
\includegraphics[width=\linewidth]{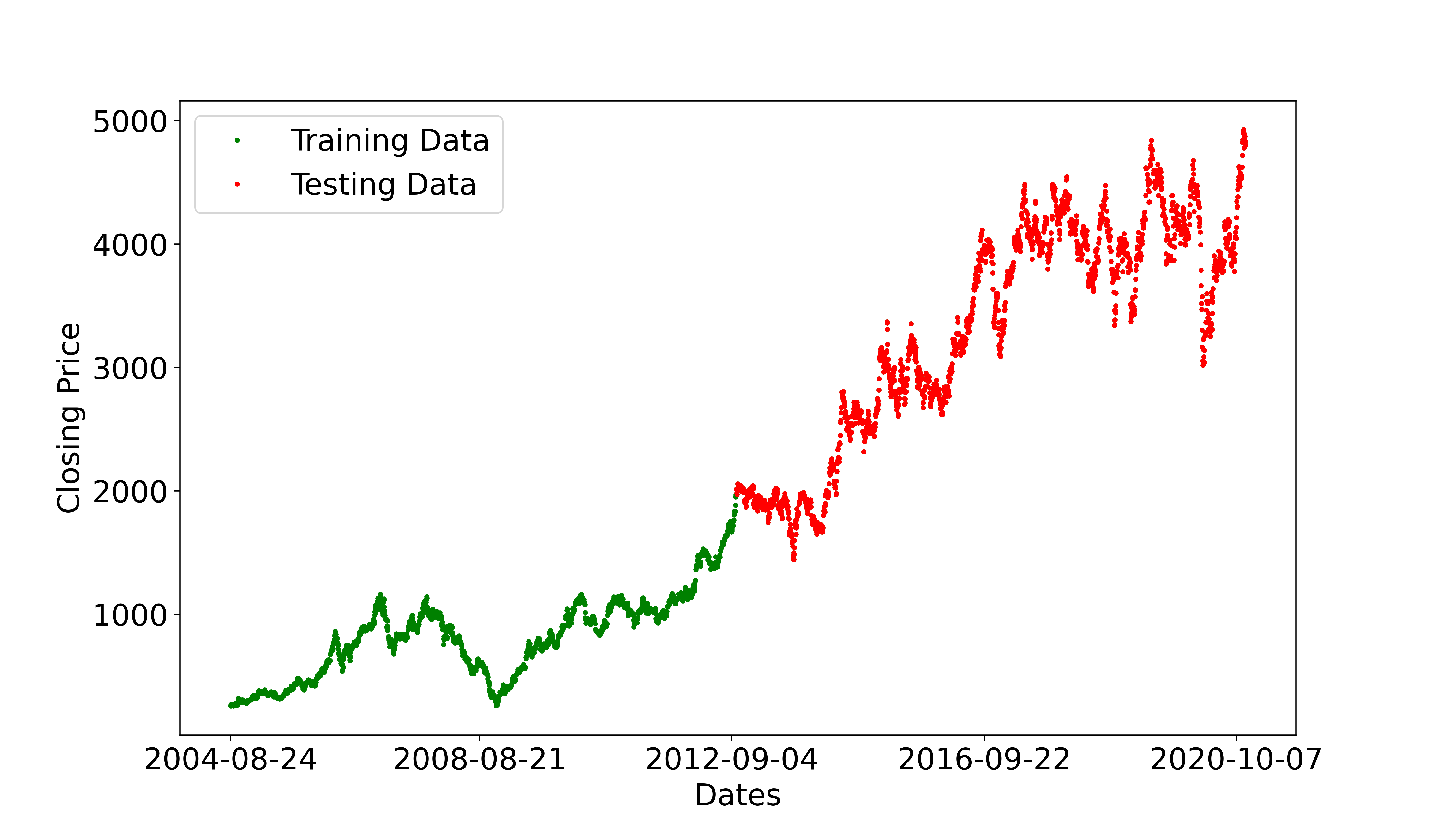}

\caption{Plot showing train and test dataset of ULTRACEMCO stock price.}
\label{fig:2}
\end{figure}

\subsection{Agent Training}
The Q-Network has input, hidden and output layers and the hyperparameters are tuned ~\ref{tab:1} to obtain the optimal weights. Tuning the hyperparameters of the model in time-series problems is very crucial for the long-term reward. The Q-Network is trained by minimizing the loss function as follows:
\begin{equation}
    L(\theta)=E[(Q_{target}-Q(s_t,a_t;\theta))^2]
\end{equation}
The learning
rate is $0.00025$ and the optimizer is Adam optimizer. The training is done for 50 episodes with batch size of 64 and the agent performs three actions: hold, buy and sell.

\subsection{Agent Testing}
The testing of the agent is done on the unseen test dataset of later periods of the same time series as the train dataset. The performance of the agent is measured in terms of total profit. The profit is calculated by sale price - purchase price.

\begin{table}
\caption{\label{tab:1}%
Model hyperparameters
}
\begin{center}
\begin{tabular}{ |c|c|c|c| } 
\hline
\bf Hyperparameters & \bf Values \\ 
\hline
Window size   & 90 \\[1ex]
\hline
Batch size   & 64 \\[1ex]
\hline
Episodes   & 50 \\[1ex]
\hline
Gamma & 0.95 \\[1ex] 
\hline
Epsilon & 1 \\[1ex] 
\hline
Learning rate & 0.00025 \\[1ex]
\hline
Epsilon minimum & 0.1 \\[1ex]
\hline
Epsilon decay & 0.995 \\[1ex]
\hline
Optimizer & Adam \\[1ex]
\hline
Loss function & Mean square error \\[1ex]
\hline
\end{tabular}
\end{center}
\end{table}
 
\section{Results}
\begin{figure*}
    \centering 
\begin{subfigure}{0.5\textwidth}
  \includegraphics[width=\linewidth]{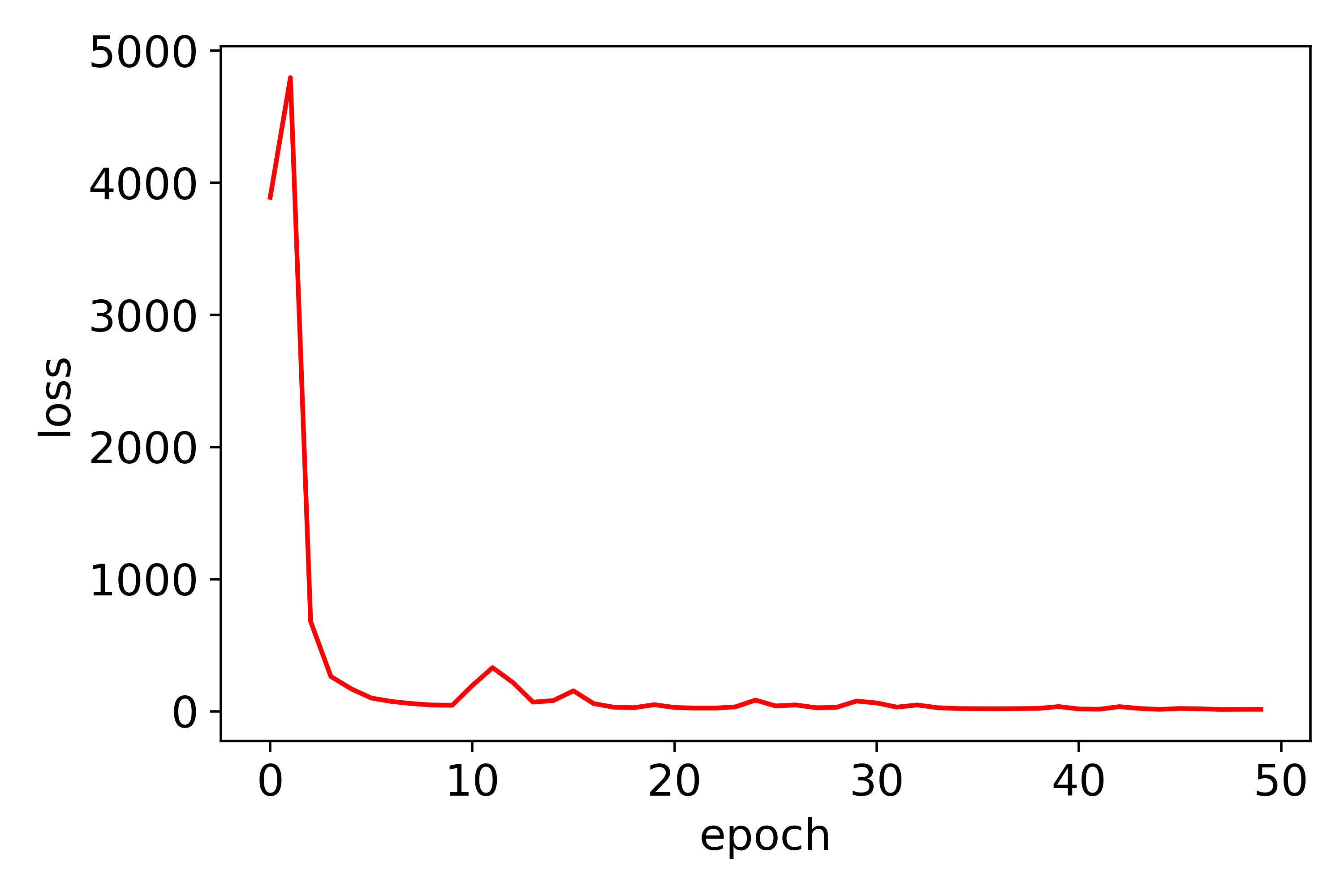}
  \caption{}
  \label{fig:3a}
\end{subfigure}\hfil 
\begin{subfigure}{0.5\textwidth}
  \includegraphics[width=\linewidth]{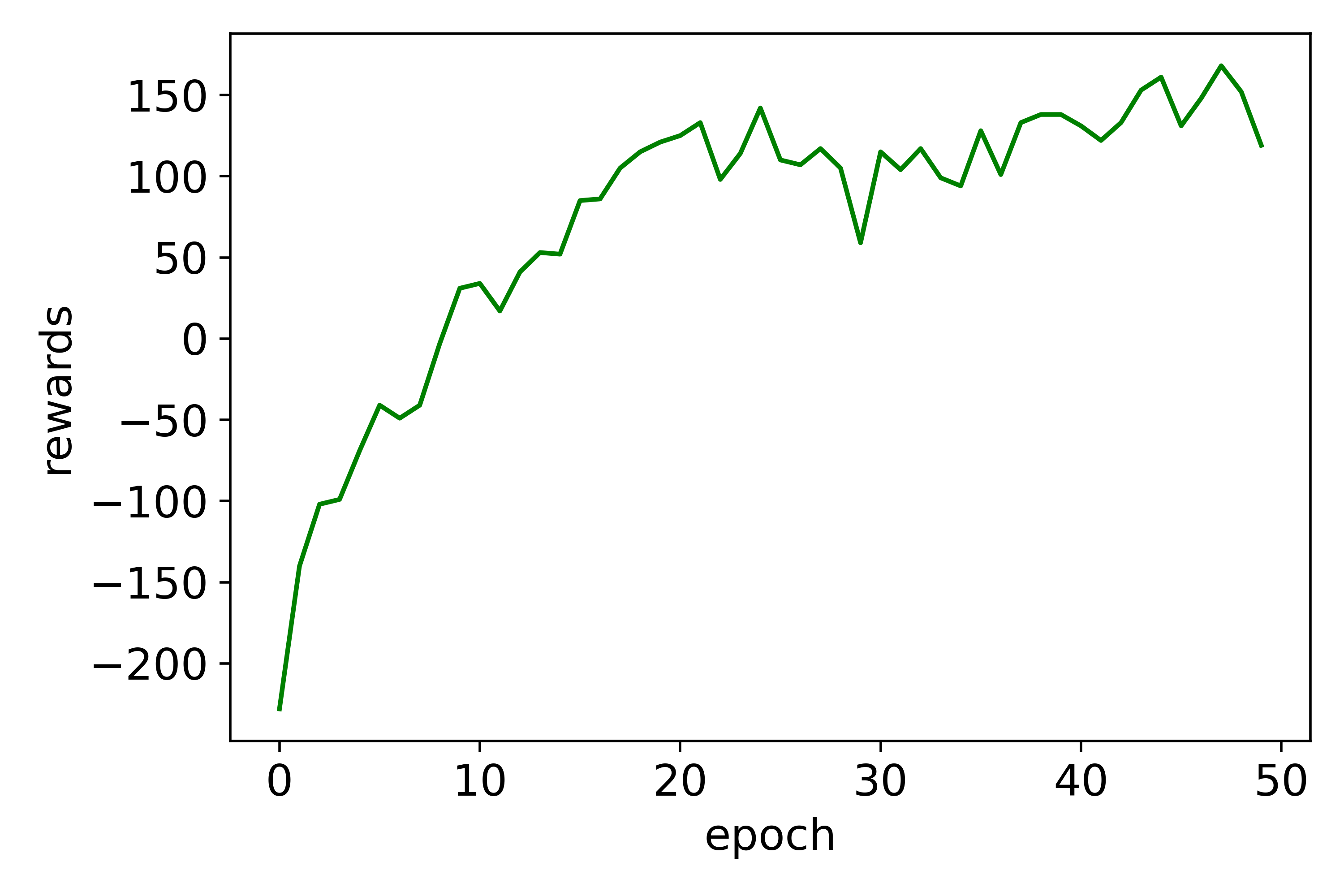}
  \caption{}
  \label{fig:3b}
\end{subfigure}\hfil 

\medskip

\begin{subfigure}{1.0\textwidth}
  \includegraphics[width=\linewidth]{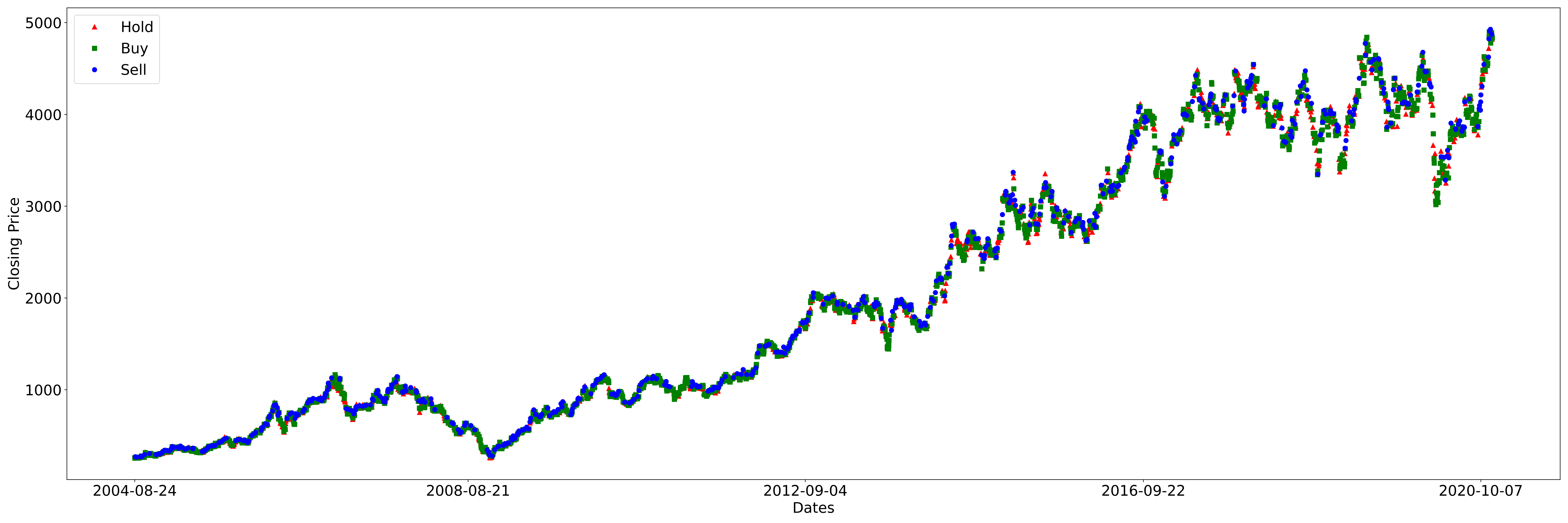}
  \caption{}
  \label{fig:3c}
\end{subfigure}\hfil 

\caption{Plots showing (a) train loss (b) train rewards (c) time-market profile of ULTRACEMCO stock using DQN  }
  \label{fig:3}
\end{figure*}

\begin{figure*}
    \centering 
\begin{subfigure}{0.5\textwidth}
  \includegraphics[width=\linewidth]{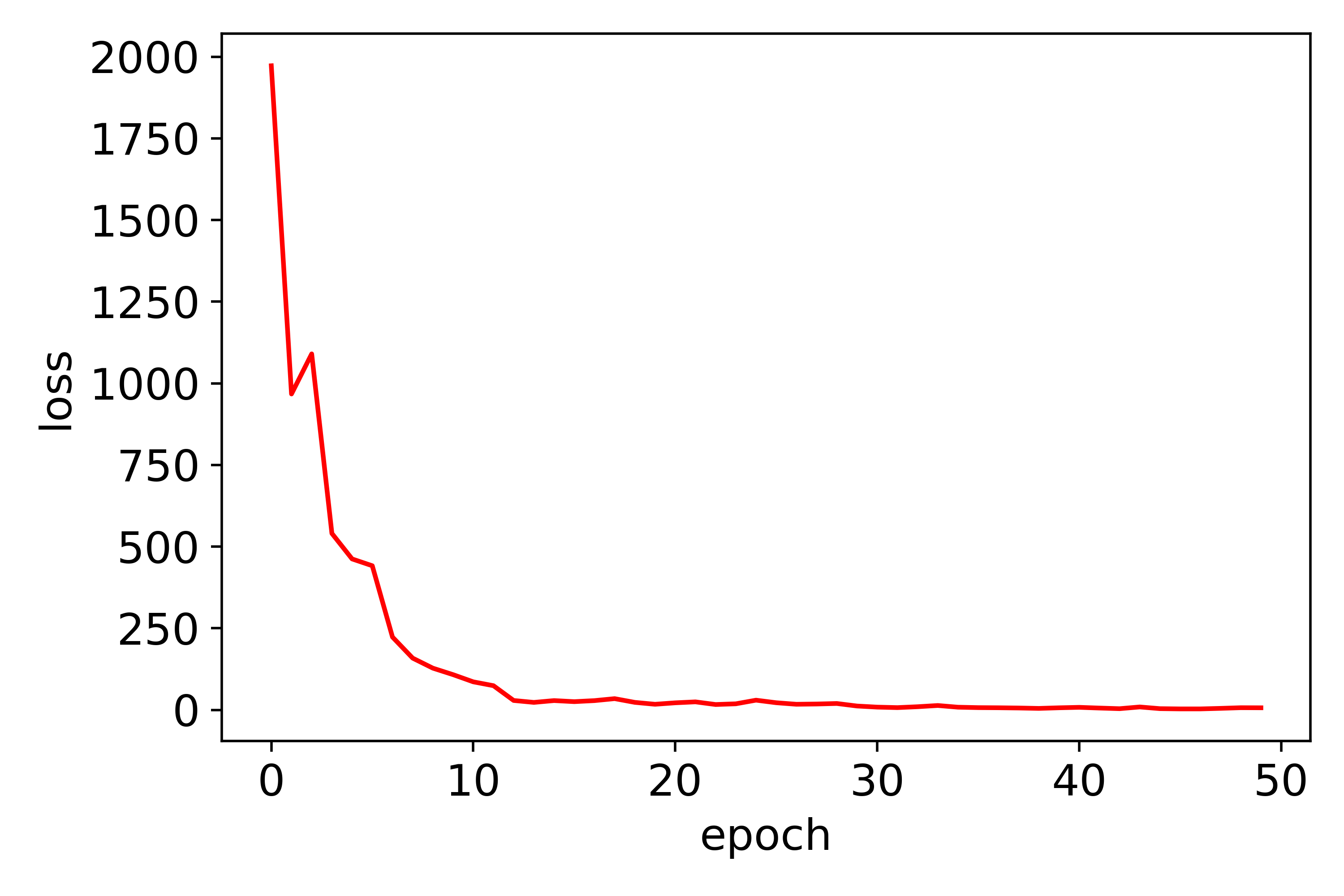}
  \caption{}
  \label{fig:4a}
\end{subfigure}\hfil 
\begin{subfigure}{0.5\textwidth}
  \includegraphics[width=\linewidth]{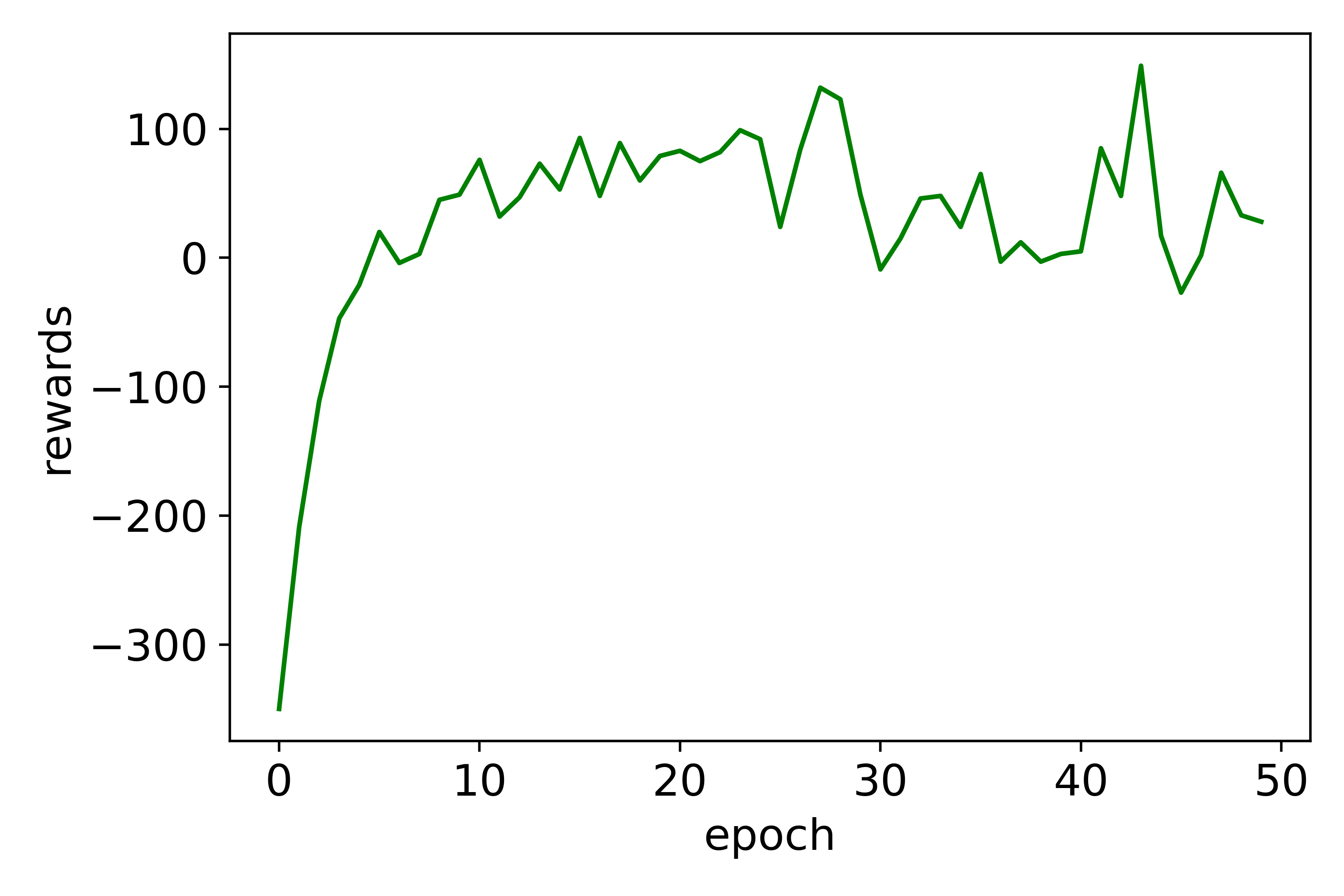}
  \caption{}
  \label{fig:4b}
\end{subfigure}\hfil 

\medskip

\begin{subfigure}{1.0\textwidth}
  \includegraphics[width=\linewidth]{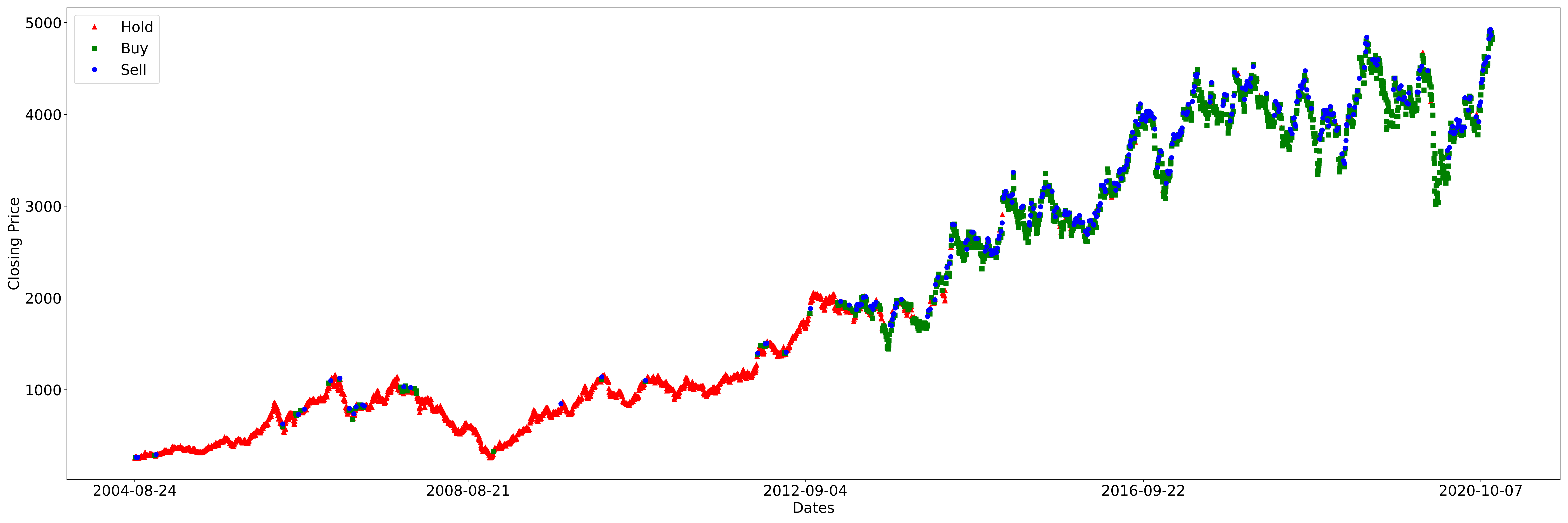}
  \caption{NESTLEIND train and test data}
  \label{fig:4c}
\end{subfigure}\hfil 

\caption{Plots showing (a) train loss (b) train rewards (c) time-market profile of ULTRACEMCO stock using Double DQN}
  \label{fig:4}
\end{figure*}

\begin{figure*}
    \centering 
\begin{subfigure}{0.5\textwidth}
  \includegraphics[width=\linewidth]{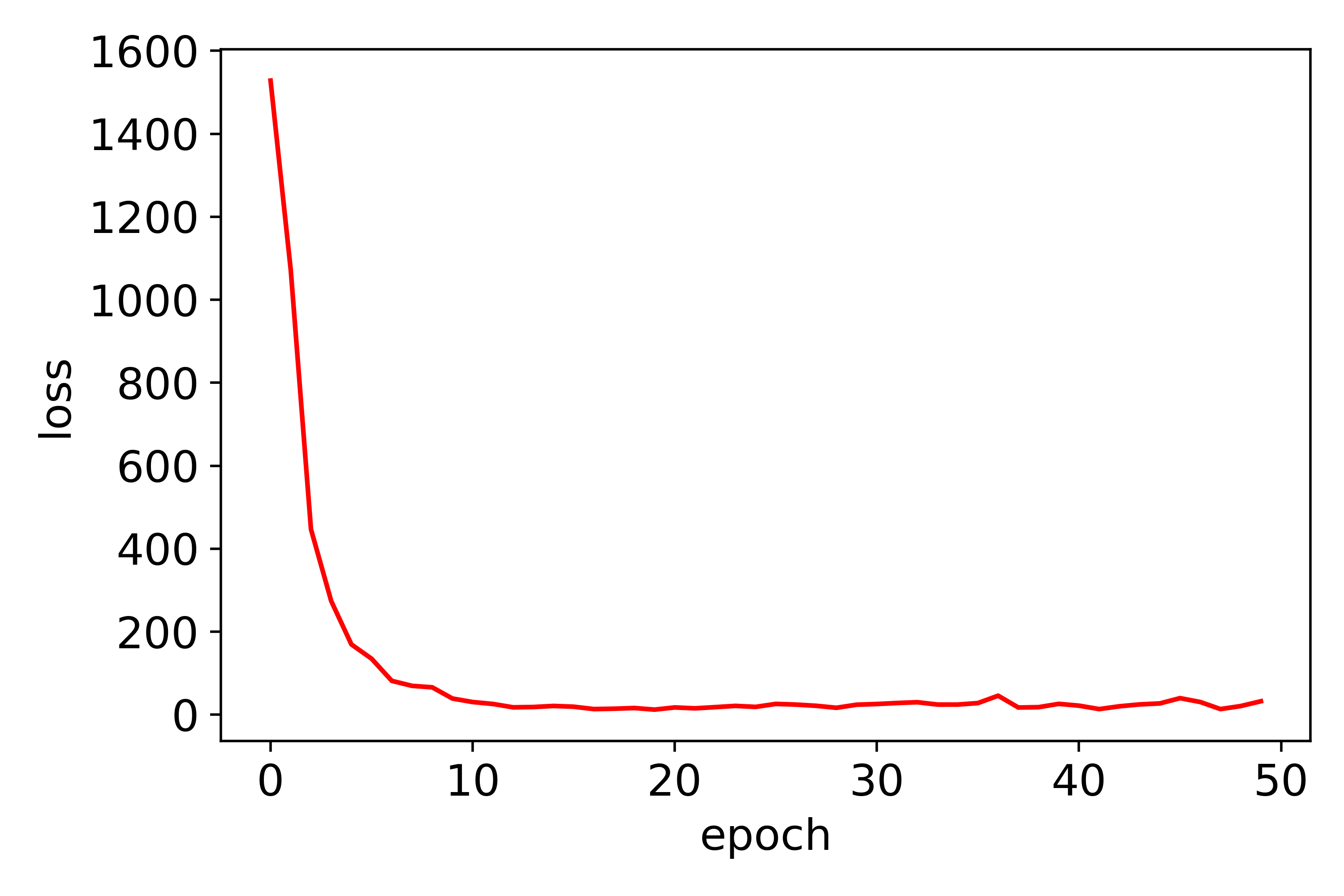}
  \caption{}
  \label{fig:5a}
\end{subfigure}\hfil 
\begin{subfigure}{0.5\textwidth}
  \includegraphics[width=\linewidth]{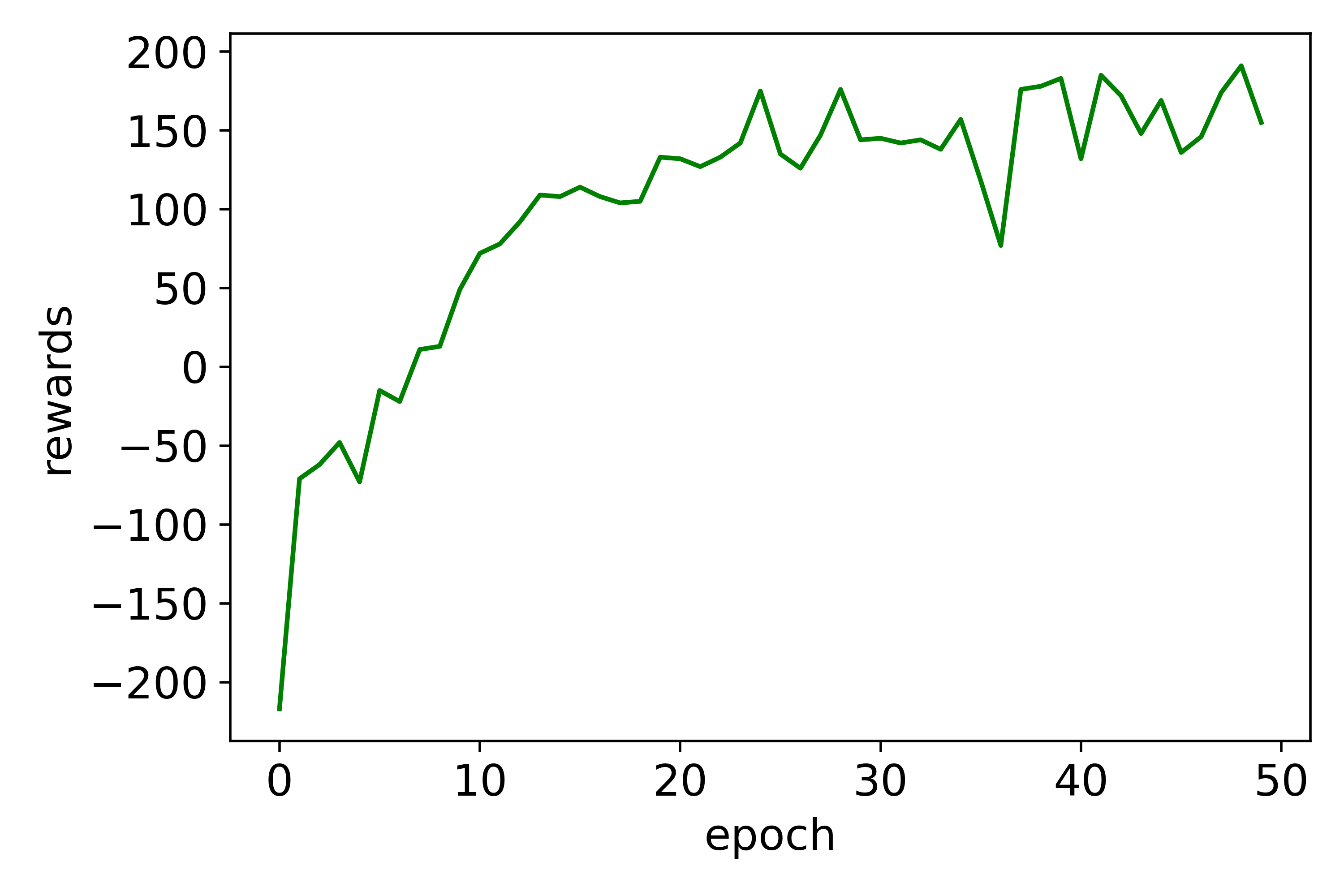}
  \caption{}
  \label{fig:5b}
\end{subfigure}\hfil 

\medskip

\begin{subfigure}{1.0\textwidth}
  \includegraphics[width=\linewidth]{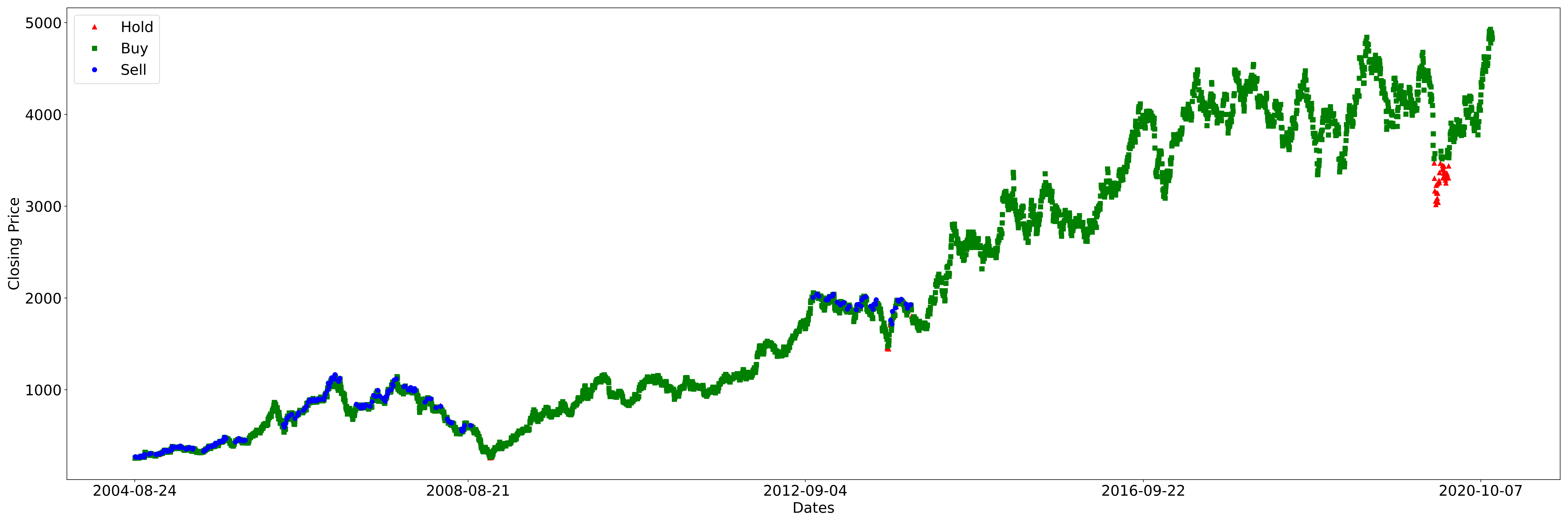}
  \caption{}
  \label{fig:5c}
\end{subfigure}\hfil 

\caption{Plots showing (a) train loss (b) train rewards (c) time-market profile of ULTRACEMCO stock using Dueling DDQN  }
  \label{fig:5}
\end{figure*}

Ten Indian stock datasets and three deep Q-networds are used to perform the experiments. Each dataset is trained on train data and tested on the unseen test data. Total rewards and profit of training data and test data is calculated for ten Indian stocks using three deep reinforcement
learning models (DQN, Double DQN and Dueling DDQN) are shown in Table~\ref{tab:2},3,4 respectively.
 Figure~\ref{fig:2} shows the train and test data used for each dataset. We randomly choose one stock dataset (ULTRACEMCO dataset) and plot the train and test data and also the training loss and training rewards with respect to number of epochs for DQN (Figure 3a,b). Mean square error is used to calculate the loss that estimates the difference between the actual and predicted values. Figure~\ref{fig:3c} shows the time-market value of the DQN model corresponding to the ULTRACEMCO dataset. Red, green and blue points corresponds to hold, buy and sell the stock respectively. Similarly, Figure~\ref{fig:4}a,b,c shows the training loss, training rewards and time-market value for the ULTRACEMCO dataset using Double DQN. Figure~\ref{fig:5}a,b,c shows the training loss, training rewards and time-market value for the ULTRACEMCO dataset using Dueling Double DQN. From Table~\ref{tab:2},3,4 we observe that on an average the  Dueling DDQN performs better than rest two models and the performance of DDQN is better than DQN.


\begin{table}[!t]
\begin{center}
\caption{\label{tab:2}%
Rewards and profit obtained during training and testing of the Indian stock datasets using DQN.}

\begin{tabular}{c c c c c c}
\hline 
\noalign{\vskip 0.5mm}
  &&  & DQN &  &  \\
\cline{2-6}
 Dataset && Train Rewards & Train Profit & Test Rewards & Test Profit \\
\hline
TCS && 246 & 12382  & 22  & 4770  \\
 \hline
RELIANCE && 117 & 17103 & -77 & 1246 \\
 \hline
ZEEL && 295 & 6639 & 124 & 2923 \\
 \hline
TATAMOTORS && 210 & 10506 & -1 & 1670 \\
\hline
TECHM  && -426 & 66 & -409 & -678  \\
\hline
UPL && 179  & 3671  & 82  & 4828  \\
 \hline
ULTRACEMCO && 199  & 8818 & 16 & 25188  \\
 \hline
TATASTEEL && 225 & 3481  & 36  & 48  \\
 \hline
NESTLEIND && -120 & 11774  & -180 & 16389  \\
 \hline
POWERGRID && 199  & 1145 & 51 & 807 \\

\hline
\end{tabular}
\end{center}
\end{table}

\begin{table}[!t]
\begin{center}
\caption{\label{tab:3}%
Rewards and profit obtained during training and testing of the Indian stock datasets using Double DQN.}

\begin{tabular}{c c c c c c}
\hline 
\noalign{\vskip 0.5mm}
  &&  & Double DQN &  &  \\
\cline{2-6}
 Dataset && Train Rewards & Train Profit & Test Rewards & Test Profit \\
\hline
TCS && 225 & 14946  & 276  & 38095  \\
 \hline
RELIANCE && -175 & 0 & -211  & 48 \\
 \hline
ZEEL && -1 & 17 & 3 & 12 \\
 \hline
TATAMOTORS && 52 & 718 & 85  & 1067  \\
\hline
TECHM  && -15 & 52  & 3  & 117 \\
\hline
UPL && 6  & 409 & 6 & 658  \\
 \hline
ULTRACEMCO && 23 & 655 & 319 & 57626  \\
 \hline
TATASTEEL && 36 & 1158 & -8 & 8  \\
 \hline
NESTLEIND && 7 & 8589 & 8 & 22016  \\
 \hline
POWERGRID && 169 & -174  & 167  & 814 \\

\hline
\end{tabular}
\end{center}
\end{table}

\begin{table}[!t]
\begin{center}
\caption{\label{tab:4}%
Rewards and profit obtained during training and testing of the Indian stock datasets using Dueling DDQN.}

\begin{tabular}{c c c c c c}
\hline 
\noalign{\vskip 0.5mm}
  &&  & Dueling DDQN &  &  \\
\cline{2-6}
 Dataset && Train Rewards & Train Profit & Test Rewards & Test Profit \\
\hline
TCS && 47 & 3497 & 114  & 17278  \\
 \hline
RELIANCE && 361 & 29392  & 347  & 29769 \\
 \hline
ZEEL && 28  & 1701 & 151 & 2836 \\
 \hline
TATAMOTORS && 250 & 16592 & 188 & 8312 \\
\hline
TECHM  && 64  & 26024 & 86 & 14831  \\
\hline
UPL && 104 & 7972 & 176 & 10284  \\
 \hline
ULTRACEMCO && 123  & 7113 & 35 & 6257 \\
 \hline
TATASTEEL && 1 & 17 & 3 & 57 \\
 \hline
NESTLEIND && 139 & 43900 & 79 & 101731  \\
 \hline
POWERGRID && 59 & 560 & 102  & 1252  \\

\hline
\end{tabular}
\end{center}
\end{table}

\section{Conclusion}
We implemented  deep reinforcement learning to automate trade execution
and generate profit. We also showed how well DRL performs in solving  stock market strategy problems and compared three DRL networks: DQN, DDQL and Dueling DDQN for 10 Indian sock datasets. The experiments showed that all these three deep learning algorithms perform well in solving the decision-making problems of stock market strategies. Since, the stock markets are highly stochastic and changes very fast, these algorithms respond to these changes quickly and perform better than traditional methods. We observe that on an average the Dueling DDQN network performed better than DDQN and DQN and Double DQN performed better than DQN.

 






\end{document}